# PARAMETRIC OSCILLATOR OF ROBUST CHAOS: CIRCUIT IMPLEMENTATION AND SIMULATION USING THE PROGRAM PRODUCT MULTISIM [1]

## S.P. Kuznetsov

Kotel'nikov's Institute of Radio-Engineering and Electronics of RAS, Saratov Branch,
Zelenaya 38, Saratov, 410019, Russian Federation

Saratov State University,
Astrakhanskaya 83, Saratov, 410012, Russian Federation

A scheme is suggested of the parametric generator of chaotic oscillations with attractor represented by a kind of Smale-Williams solenoid that operates under a periodic sequence of pump pulses at two different frequencies. Simulation of chaotic dynamics using the software product Multisim is provided.

**Introduction**

Parametric oscillations are known in mechanics, electronics, acoustics, nonlinear optics [1-7]. One of the most popular examples relates to swinging when a person gradually increases the amplitude of the oscillatory motion by changing the position of the body, which corresponds to the periodic variation of the effective length of the equivalent pendulum. One more example is an electrical resonant circuit in which the amplitude of the oscillations can be increased step by step by periodically varying capacitance of the capacitor: increasing distance between the plates at times corresponding to the maximum charge, and returning them to the original position at the instants of maximal current in the inductor, when the capacitor charge is close to zero.

The principle of parametric excitation of oscillations seems promising from the point of view of an interesting problem of design of physical devices with chaotic dynamics associated with so-called uniformly hyperbolic attractors [8-11]. These objects in state space of dynamical systems are composed exclusively of saddle trajectories, combining instability of the orbits on the attractor with stability in the sense of approach of transient trajectories to the attractor. As abstract artificial objects, such attractors (Smale-Williams solenoid, Plykin attractor, DA-attractor of Smale) have been introduced in the mathematical theory of dynamical systems about half a century ago, but until recently they have not been considered in the context of natural science and technical applications. Such a situation is odd and unacceptable; indeed, the uniformly hyperbolic attractors form a unique class of chaotic attractors, for which the property of roughness, or structural stability, is mathematically proven. It guarantees that the chaotic dynamics will be insensitive to variations of functions and parameters in the respective dynamical equations. As postulated in the theory of oscillations, roughness is a fundamental basis for significance in practical applications and for priority theoretical analysis of the systems [12]. For any possible practical use (secure communication [13], noise radar [14], generation of random numbers [15]) it will be natural to give preference to the generators of rough chaos.

Consider geometric construction of the Smale-Williams attractor essential for further discussion. For clarity, the explanation relates to the case of a discrete time system with three-dimensional state space. Suppose that at one step of evolution some domain in a form of a torus undergoes strong transversal compression and triple longitudinal expansion being then folded to a triple loop and located inside the original torus (Figure 1, panel (a)). At each step of the transformation, the full volume of the object is reduced (dissipative system), and the number of loops is tripled. In the limit of infinite number of steps the number of coils tends to infinity, and the solenoid appears, which has a structure of Cantor set in the transversal direction (panel (b)).

The most important point is that the angular coordinate measured along the solenoid coils is governed by a map with chaotic dynamics. In our example, this is a three-fold expanding circle map $\varphi_{n+1} = -3\varphi_n + \text{const}$ (a constant is determined by selection of the origin of the variable $\varphi$). Iteration

---

diagram for this map is shown in panel (c) of Fig.1. The characteristic feature of chaos is instability to perturbations of the initial conditions. In the present case, a small disturbance of the variable φ is tripled on one step that corresponds to a positive Lyapunov exponent $\ln 3 = 1.0986...$

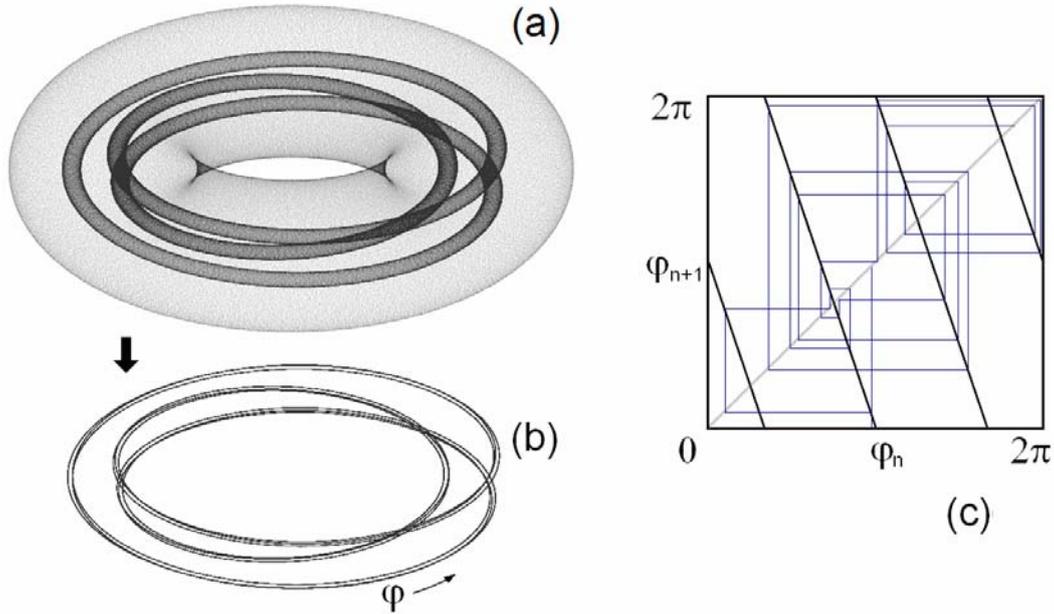

**Fig.1.** Transformation of a domain of toroidal shape by one step of evolution in discrete time (a) and the solenoid attractor appearing in the limit of a large number of steps (b). The tripling map for angular coordinate and its chaotic behavior are illustrated by diagram (c).

This example is a typical construction offered by mathematicians who appeal in their considerations to geometrical, topological, algebraic concepts. Alternatively, a physicist should apply his own toolbox to construct models with structurally stable chaos involving oscillators, non-linear elements, feedback loops etc.

A general principle of designing physical systems with hyperbolic attractors proposed and demonstrated in [16-21] consists in manipulating with phases in the course of transmission of excitation between two or more alternately active partial oscillators, so that the respective map for the phases would exhibit chaotic dynamics.

Parametric generator of chaos on this principle described in [22] contains two identical subsystems, each of which consists of two parametrically coupled oscillatory circuits with natural frequencies ω and 2ω. An oscillator of frequency ω in one subsystem is connected via a quadratic nonlinear element with an oscillator of frequency 2ω of other subsystem. Pump at frequency 3ω is supplied to the subsystems alternately. In Ref. [23] another scheme is proposed that uses only one parametric oscillator with modulated pump supplemented with a delayed feedback loop containing a quadratic nonlinear element. One more scheme [24] is based on two parametrically coupled oscillators, where the natural frequencies are in ratio 1:2, and pump is produced at the triple frequency by periodic pulses accompanied by Q-switching with the same period. In all these systems attractors of Smale – Williams type occur because the angular variable associated with internal phase of successively generated oscillation trains corresponds to the expanding circle map.

All these schemes are characterized by a fixed ratio of frequencies of oscillators and the pump 1: 2: 3, and, apparently, do not use all possibilities of manipulation by frequencies and phases available with the parametric excitation. Moreover, if one talks about design of concrete electronic devices the problem needs more deep elaboration. Say, one should take into account that the varactor diode as the most applicable non-linear element in construction of electronic parametric oscillators has the nonlinear characteristic (capacitance versus voltage) not reducible to simple quadratic or cubic function, and requires suitably selected bias for proper operation.

In this paper we propose a parametric generator of hyperbolic chaos with attractor of Smale – Williams type, wherein the frequency ratio can be set with a high degree of arbitrariness, pumping is supplied alternately at two frequencies, the sum and the difference of the natural frequencies of the oscillators, and Q-switching is not used at all. The principle of operation will be illustrated for model equations, then it is implemented in an electronic circuit with varactor diodes, and its functioning is demonstrated by simulation of the dynamics by means of the software product Multisim.

Although some electronic circuits with uniformly hyperbolic attractors of Smale-Williams type, Plykin attractor, and DA-attractor were already published [25-28], they did not belong to the class of parametric oscillators; moreover, they required more complex element base (operational amplifiers, multipliers, etc.).

**1. The model equations and the operation principle of the system**

Let's start with introductory explanations of the active and passive parametric coupling.

Consider two oscillators with natural frequencies $\omega_1$ and $\omega_2$ (so far without dissipation) and introduce reactive coupling oscillating with the pumping frequency $\Omega$ and characterized by a coefficient $\varepsilon$:

$$\ddot{x}_1 + \omega_1^2 x_1 = \varepsilon x_2 \sin \Omega t,$$
$$\ddot{x}_2 + \omega_2^2 x_2 = \varepsilon x_1 \sin \Omega t. \qquad (1)$$

If the frequency $\Omega$ is equal to the sum of the natural frequencies of the partial oscillators, the parametric instability occurs that is a simultaneous growth of the oscillations in both subsystems; the energy is supplied from the pump source. If we take the value $\Omega$ equal to the difference of the natural frequencies, then the passive parametric interaction takes place, which manifests itself in beats, slow alternate growth and decrease of amplitudes of one or the other oscillator; the energy is periodically transferred to each other.

For an approximate analytical description of these effects a method of slow amplitudes is appropriate. Instead of the coordinates and velocities, we introduce complex variables $a_k(t)$ setting

$$x_k = a_k(t) e^{i\omega_k t} + a_k^*(t) e^{-i\omega_k t}, \quad \dot{x}_k = i\omega_k a_k(t) e^{i\omega_k t} - i\omega_k(t) a_k^* e^{-i\omega_k t}, \qquad (2)$$

where $k=1,2$. As evident from (2), the new variables must obey the relations

$$\dot{a}_k(t) e^{i\omega_k t} + \dot{a}_k^*(t) e^{-i\omega_k t} = 0. \qquad (3)$$

Now, we rewrite the equation (1) in new variables by the substitution (2) and exclude the conjugate derivatives $\dot{a}_{1,2}^*$ with the help of (3). Next, we perform averaging over a period of fast oscillations that is time scale, on which the complex amplitudes $a_{1,2}$ can be considered roughly constant.

In the case $\Omega = \omega_2 - \omega_1$ we have

$$\dot{a}_1 = \tfrac{1}{4}\varepsilon\omega_1^{-1} a_2, \quad \dot{a}_2 = -\tfrac{1}{4}\varepsilon\omega_2^{-1} a_1, \qquad (4)$$

and get the oscillatory solution $a_1 \sim \sin[\tfrac{1}{4}\varepsilon(\omega_1\omega_2)^{-1/2} t]$, $a_2 \sim \cos[\tfrac{1}{4}\varepsilon(\omega_1\omega_2)^{-1/2} t]$ that corresponds to the frequency of beats $\tfrac{1}{4}\varepsilon(\omega_1\omega_2)^{-1/2}$.

On the other hand, in the case $\Omega = \omega_2 + \omega_1$ the equations take the form

$$\dot{a}_1 = -\tfrac{1}{4}\varepsilon\omega_1^{-1} a_2^*, \quad \dot{a}_2 = -\tfrac{1}{4}\varepsilon\omega_2^{-1} a_1^*. \qquad (5)$$

As follows, $\ddot{a}_{1,2} = \tfrac{1}{16}\varepsilon^2(\omega_1\omega_2)^{-1} a_{1,2}$, and we get the exponentially growing amplitudes in both subsystems $a_{1,2} \sim \exp[\tfrac{1}{4}\varepsilon(\omega_1\omega_2)^{-1/2} t]$. To be physically justified, the model for this case should take into account saturation of the amplitude; for this the equations must be complemented with additional terms introducing nonlinear dissipation or nonlinear offset of the oscillation frequency.

Now, let us consider a system of three oscillators. Assume that the first and the third oscillators interact through reactive element with cubic nonlinearity and their natural frequencies satisfy a condition $\omega_3 = 3\omega_1$ while the frequency of the second oscillator $\omega_2$ is substantially greater. Pumping is provided by pulses of frequency $\omega_1 + \omega_2$ and $\omega_2 - \omega_3$ applied alternately, so that the operation of the system consists in repetition of three stages of time duration $T_1$, $T_2$, and $T_3$, with the period $T = T_1 + T_2 + T_3$. The system is supposed to be dissipative.

The model equations are [2]

$$\ddot{x}_1 + \omega_1^2 x_1 = -\alpha_1 \dot{x}_1 + 3\varepsilon_{13} x_1^2 x_3 + \varepsilon_{12} x_2 f(t) \sin(\omega_1 + \omega_2)t,$$
$$\ddot{x}_2 + \omega_2^2 x_2 = -\alpha_2 \dot{x}_2 - \beta \dot{x}_2^3 + \varepsilon_{12} x_1 f(t) \sin(\omega_1 + \omega_2)t + \varepsilon_{23} x_3 g(t) \sin(\omega_2 - 3\omega_1)t, \qquad (6)$$
$$\ddot{x}_3 + \omega_3^2 x_3 = -\alpha_3 \dot{x}_3 + \varepsilon_{13} x_1^3 + \varepsilon_{23} x_2 g(t) \sin(\omega_2 - 3\omega_1)t,$$

where

$$f(t) = \begin{cases} 1, & 0 \le t < T_1, \\ 0, & T_1 \le t < T, \end{cases} \quad g(t) = \begin{cases} 0, & 0 \le t < T - T_3, \\ 1, & T - T_3 \le t < T, \end{cases} \quad f(t+T) = f(t),\ g(t+T) = g(t).$$

To simplify mathematical description of the problem, we assume that duration of each stage contains an integer number of periods of oscillation signals of the pump; namely, $(\omega_1 + \omega_2)T_1/2\pi$ and $(\omega_2 - 3\omega_1)T_3/2\pi$ are integer numbers. So, the equations (6) have time-periodic coefficients. For this non-autonomous system one can consider dynamics in discrete time by means of the Poincaré map. In this approach we track the states of the system stroboscopically, at time instants $nT$. The Poincaré map $\mathbf{x}_{n+1} = \mathbf{T}(\mathbf{x}_n)$ is defined as a transformation of the six-dimensional state vector $\mathbf{x} = \{x, \dot{x}, y, \dot{y}, z, \dot{z}\}$ on the modulation period of the pump. Practically, the Poincaré map is carried out by a computer program that performs numerical solution of the equations (6) on a time interval equal to the period $T$.

In the first stage of the system functioning, for a time interval $T_1$, the pumping of intensity determined by the coefficient $\varepsilon_{12}$ at the frequency $\Omega_1 = \omega_1 + \omega_2$ provides parametric excitation of the first and the second oscillators. The oscillations are characterized by some phase constant $\varphi$ determined by initial conditions in the beginning of the stage: $x_1 \sim \sin(\omega_1 t + \varphi)$, $x_2 \sim \sin(\omega_2 t - \varphi)$. Saturation of the amplitude of the oscillations is achieved due to nonlinear dissipation in the second oscillator characterized by parameter $\beta$.

In the second stage of duration $T_2$ the pump is switched off, and damping occurs due to the dissipation parameters $\alpha_{1,2}$. Because of the coupling through the cubic nonlinearity between the first and the third oscillator with the resonance frequency ratio 1: 3, which is characterized by the parameter $\varepsilon_{13}$, the third oscillator with a relatively small damping parameter $\alpha_3$ undergoes the amplitude growth, and the resulting phase of the induced oscillations corresponds to the phase of the third harmonic of the first oscillator: $x_3 \sim \sin(3\omega_1 t + 3\varphi + \text{const}) = \sin(\omega_3 t + 3\varphi + \text{const})$.

In the third stage the passive parametric coupling is provided by oscillation of the coupling parameter of the second and the third oscillators at the frequency $\Omega_2 = \omega_2 - \omega_3$ with magnitude $\varepsilon_{23}$. With properly selected duration of the stage $T_3$, due to the beats, a practically complete transfer of energy from the third to the second oscillator will take place, and the second oscillator eventually inherits the phase from the third oscillator: $x_2 \sim \sin(\omega_2 t + 3\varphi + \text{const})$. Thus, to the beginning of the next stage of parametric excitation, when the pumping at frequency $\Omega_1$ is switched on, the updated value of the phase constant is given by the tripled initial value with the opposite sign: $\varphi_{new} = -3\varphi + \text{const}$. Then the process is repeated periodically.

---

[2] The model is intended only to demonstrate the principle of operation of the proposed scheme, but does not pretend for quantitative description of the electronic device discussed in Section 3. Nevertheless, evident qualitative agreement gives a foundation to talk about occurrence of the hyperbolic Smale-Williams attractor in the electronic device accounting structural stability of the phenomenon.

The tripling of the angular variable φ accompanying by compression of the phase volume due to dissipation in other directions of the state space just corresponds to the situation of formation of the Smale-Williams attractor as discussed in the Introduction.

One can apply the method of slow amplitudes to the equations (6) using relations (2) and (3), where the index *k* takes values from 1 to 3. In analogy to derivations in the beginning of this Section, it is possible to write out the complex amplitude equations for each stage as follows.

Stage I, $nT \leq t < nT + T_1$:

$$\dot{a}_1 + \tfrac{1}{2}\alpha_1 a_1 + \tfrac{3}{2}i\varepsilon_{13}\omega_1^{-1}a_1^{*2}a_3 = -\tfrac{1}{4}\varepsilon_{12}\omega_1^{-1}a_2^*,$$
$$\dot{a}_2 + \tfrac{1}{2}\alpha_2 a_2 + \tfrac{3}{2}\omega_2^2\beta |a_2|^2 a_2 = -\tfrac{1}{4}\varepsilon_{12}\omega_2^{-1}a_1^*, \quad (7a)$$
$$\dot{a}_3 + \tfrac{1}{2}\alpha_3 a_3 + \tfrac{1}{2}i\varepsilon_{13}\omega_3^{-1}a_1^3 = 0.$$

Stage II, $nT + T_1 \leq t < nT + T_2$:

$$\dot{a}_1 + \tfrac{1}{2}\alpha_1 a_1 + \tfrac{3}{2}i\varepsilon_{13}\omega_1^{-1}a_1^{*2}a_3 = 0,$$
$$\dot{a}_2 + \tfrac{1}{2}\alpha_2 a_2 + \tfrac{3}{2}\omega_2^2\beta |a_2|^2 a_2 = 0, \quad (7b)$$
$$\dot{a}_3 + \tfrac{1}{2}\alpha_3 a_3 + \tfrac{1}{2}i\varepsilon_{13}\omega_3^{-1}a_1^3 = 0.$$

Stage III, $nT + T_2 \leq t < (n+1)T$:

$$\dot{a}_1 + \tfrac{1}{2}\alpha_1 a_1 + \tfrac{3}{2}i\varepsilon_{13}\omega_1^{-1}a_1^{*2}a_3 = 0,$$
$$\dot{a}_2 + \tfrac{1}{2}\alpha_2 a_2 + \tfrac{3}{2}\omega_2^2\beta |a_2|^2 a_2 = -\tfrac{1}{4}\varepsilon_{23}\omega_2^{-1}a_3, \quad (7c)$$
$$\dot{a}_3 + \tfrac{1}{2}\alpha_3 a_3 + \tfrac{1}{2}i\varepsilon_{13}\omega_3^{-1}a_1^3 = \tfrac{1}{4}\varepsilon_{23}\omega_3^{-1}a_2.$$

These equations also can be used to describe dynamics in terms of the Poincaré map that determines transformation of the state on one period of the pump modulation $\mathbf{x}_{n+1} = \mathbf{T}(\mathbf{x}_n)$, where a six-dimensional state vector is a set of real and imaginary parts of complex numbers $a_k$: $\mathbf{x} = \{\text{Re}\,a_1, \text{Im}\,a_1, \text{Re}\,a_2, \text{Im}\,a_2, \text{Re}\,a_3, \text{Im}\,a_3\}$. In the approach using slow amplitudes the above-mentioned condition of integer normalized lengths of the stages is insignificant and can be ignored.

**2. Results of numerical simulation and observation of the Smale-Williams attractor**

Numerical calculations show that the three-fold expanding circle map for the angle variable on a period of the pump modulation *T* indeed occurs in the present system.

Fig. 2 shows typical plots of time dependences of the variables *x*, *y*, *z* obtained by numerical integration of the equations (6) by a finite difference method for the parameter set $\omega_1=2\pi$, $\omega_2=10\pi$, $\omega_3=6\pi$, $T=50$, $T_1=20$, $T_3=3$, $\beta=0.008$, $\varepsilon_{12}=48$, $\varepsilon_{23}=100$, $\varepsilon_{13}=1$, $\alpha_1=0.4$, $\alpha_2=1$, $\alpha_3=0$. Chaos manifests itself in random-like variation of the oscillation phases for successive stages of activity.

Figure 3a shows a diagram for the angular variable that is the main evidence of presence of the Smale-Williams attractor for the map, which describes transformation of the system state for a period *T*. In the course of numerical integration of the differential equations, at the time instants $t_n = nT + T/2$ we determine phases $\varphi_n = \arg(x(nT + T/2) - i\omega_1^{-1}\dot{x}(nT + T/2))$ of the first oscillator and presented them graphically in coordinates $(\varphi_n, \varphi_{n+1})$. Observe that the points are definitely disposed along the branches, forming the graph of the three-fold expanding circle map. Significant is the topological nature of the transformation of the phases: one full round for pre-image $\varphi_n$ (i.e. a change of the variable by $2\pi$) corresponds to a three-fold bypass for the image $\varphi_{n+1}$ in the opposite direction. On this basis, we conclude that action of the Poincaré map, defined as the transformation of the vector **x** in the six-dimensional space on a period of the pump modulation, is accompanied by stretching the angular variable (phase) and by compression in the remaining five directions. (The presence of the expansion and compression is confirmed below by analysis of the spectrum of Lyapunov exponents.) Therefore, in the six-dimensional space one can determine a domain $D$, containing the attractor as a direct product of a one-dimensional circle and a five-dimensional ball. A single iteration of the Poincaré map in application to the domain $D$ creates an object $\mathbf{T}(D)$ in the form of a closed tube, extended in length, compressed in width and embedded as a triple folded loop within the original toroidal domain, with direction of the angular coordinate inverted compared to the original. This corresponds to the solenoid of the Smale-Williams type, described in the introduction; the only difference is that it "lives" in the six-dimensional state space of the map **T** rather than in the three-dimensional space. Portrait of the attractor in the stroboscopic section in projection onto a plane is shown in Figure 3b.

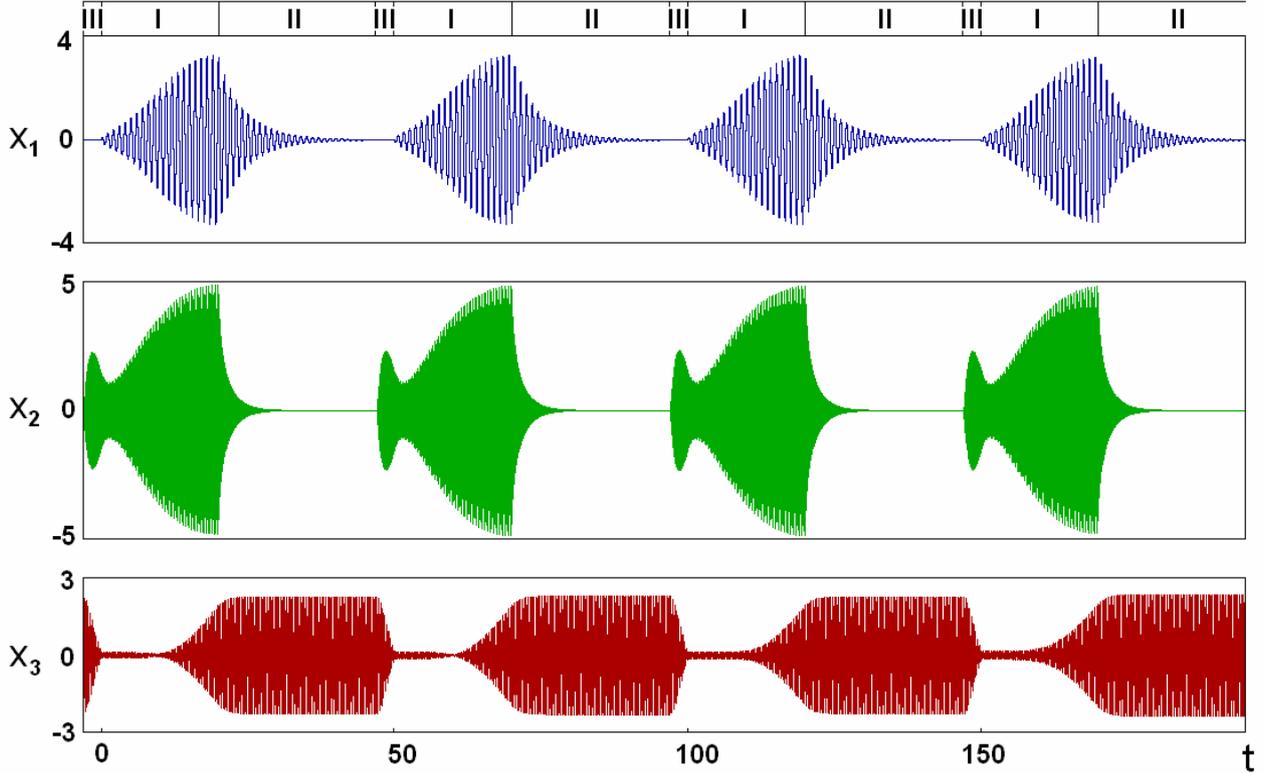

Figure 2: Time dependences for dynamic variables obtained after excluding transients in the numerical solution of equations (6) by a finite-difference method. The parameters are $\omega_1=2\pi$, $\omega_2=10\pi$, $\omega_3=6\pi$, $T=50$, $T_1=20$, $T_3=3$, $\beta=0.008$, $\varepsilon_{12}=48$, $\varepsilon_{23}=100$, $\varepsilon_{13}=1$, $\alpha_1=0.4$, $\alpha_2=1$, $\alpha_3=0$. In the upper part of the figure the Roman numerals in a ruler indicate the stages of the operation of the system.

Similar results are obtained with the slow amplitude equations. Figure 4 shows a diagram for the phase defined as the argument of the complex amplitude $a_1$ at time instants $t_n = nT + T/2$ and a portrait of the attractor in the stroboscopic section. Minor quantitative differences in comparison with Figure 3 arise apparently due to the approximate nature of description in terms of slow amplitudes.

To compute the Lyapunov exponents we follow the well-known technique [29] and undertake numerical solution of the equations (6) or (7) together with a collection of six sets of equations in variations

$$\dot{\mathbf{x}} = \mathbf{F}(\mathbf{x},t),$$
$$\dot{\tilde{\mathbf{x}}}_k = \mathbf{F}'(\mathbf{x},t)\tilde{\mathbf{x}}_k, \ k=1,2,...6. \qquad (8)$$

Here the components of vector function $\mathbf{F}(\mathbf{x},t)$ are defined by right-hand parts of the equations (6) or (7), and $\mathbf{F}'(\mathbf{x},t)$ it the matrix derivative that is a matrix of size 6x6 composed of partial derivatives of the components of **F** over the components of **x**. The tilde marks perturbation vectors tracked in the process of integration of the equations along the reference trajectory. The procedure is supplemented by normalization and orthogonalization of six perturbation vectors on each next period of the pump modulation *T*. Lyapunov exponents are obtained as the average rate of growth or decrease of the accumulated sums of logarithms of norms for the vectors which appear arranged naturally in the descending order. According to the calculations, the Lyapunov exponents for the attractor of the Poincaré map at the assigned parameters for the initial equations (6) are

$$\Lambda_1 \approx 1.093,\ \Lambda_2 \approx -1.338,\ \Lambda_3 \approx -5.64,\ \Lambda_4 \approx -7.78,\ \Lambda_5 \approx -30.6,\ \Lambda_6 \approx -32.3, \qquad (9)$$

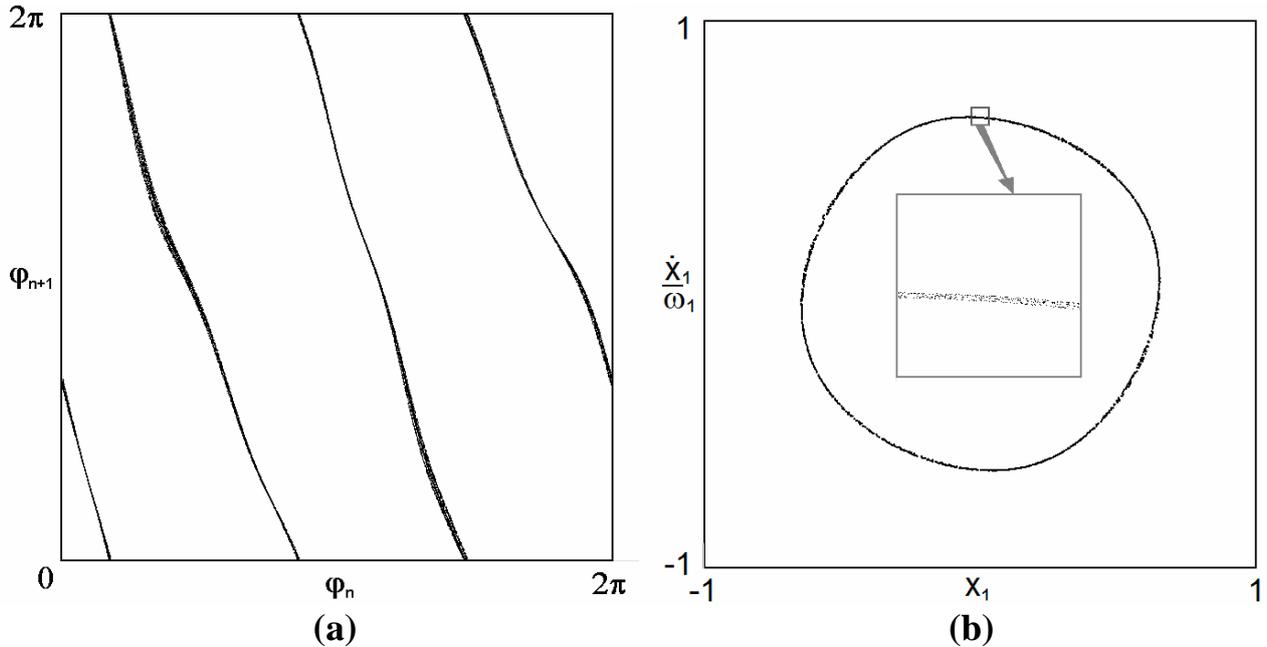

**Figure 3:** Iteration diagram for phases of oscillations of the first oscillator (a) and portrait of the attractor in the Poincaré section (b) resulting from numerical solution of the equations (6) at parameters specified in the text. Inset on the panel (b) illustrates the fine transversal structure of the filaments of the attractor.

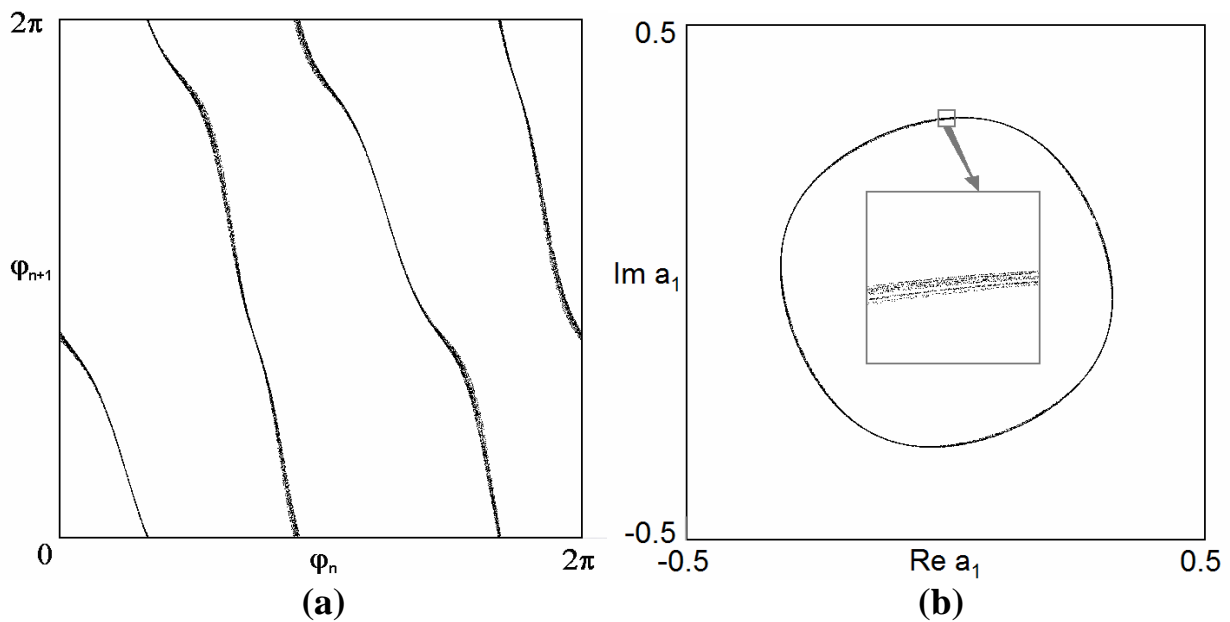

Figure 4: Iteration diagram for phases of oscillations of the first oscillator (a) and portrait of the attractor in the Poincaré section (b) resulting from numerical solution of the equations for the complex amplitudes (7) at parameters specified in the text. Inset on the panel (b) illustrates the fine transversal structure of the filaments of the attractor.

and those for the slow-amplitude equations (7) are

$$\Lambda_1 \approx 1.095, \Lambda_2 \approx -1.432, \Lambda_3 \approx -5.29, \Lambda_4 \approx -7.59, \Lambda_5 \approx -32.8, \Lambda_6 \approx -34.9. \qquad (10)$$

Presence of a positive exponent $\Lambda_1$ indicates chaotic nature of the dynamics. Its value is close to $\ln 3 = 1.0986...$ that agrees well with the approximate description of the evolution of the phase variable by the three-fold expanding circle map. Other exponents are negative and respond for approach of phase trajectories to the attractor.

It is interesting to perform direct numerical verification of hyperbolicity of the attractor at least with the slow amplitude equations (7) that requires less computing resources.

The idea of the test by the "angle criterion" was originally proposed in [30] and [31], and in the most simple and convenient version it was reformulated in [32]. In the case of one-dimensional unstable subspace (one positive Lyapunov exponent), it is as follows.

We start with evaluation of a reference orbit $\mathbf{x}(t)$ on the attractor producing numerical integration of the equation $\dot{\mathbf{x}} = \mathbf{F}(\mathbf{x},t)$ on a sufficiently long time interval. Next, we take the linearized equation for a perturbation vector $\dot{\tilde{\mathbf{x}}} = \mathbf{F}'(\mathbf{x}(t),t)\tilde{\mathbf{x}}$ and integrate it along the obtained trajectory $\mathbf{x}(t)$ with normalization of the vector $\tilde{\mathbf{x}}$ at each step of the Poincaré map $n$ to exclude divergence. It results in recording a set of unit vectors $\{\mathbf{x}_n\}$ relating to the unstable subspace. Then, according to the idea [32], we produce integration of the linear equation

$$\dot{\mathbf{u}} = -[\mathbf{F}'(\mathbf{x}(t),t)]^{\mathrm{T}} \mathbf{u}, \qquad (11)$$

along the same reference trajectory in backward time. (Here the superscript T designates the matrix conjugation.) With normalization, it yields a set of unit vectors $\{\mathbf{u}_n\}$ orthogonal to the stable subspace nearby the reference orbit. Now, to compute the angles $\alpha_n$ between the one-dimensional unstable subspace and the stable subspace (five-dimensional in our case), at each $n$-th passage of the Poincaré section we evaluate the angle $\beta_n \in [0,\pi/2]$ between $\tilde{\mathbf{x}}_n$ and $\mathbf{u}_n$ from the relation $\cos\beta_n = |\mathbf{u}_n(t) \cdot \tilde{\mathbf{x}}_n(t)|$, and set $\theta_n = \pi/2 - \beta_n$.

For an orbit long enough, processing of the accumulated data provides a statistical distribution of the angles θ. If it is definitely separated from zero, the test confirms hyperbolicity. Alternatively, if it shows non-zero probability of zero angles, it implies violation of the hyperbolicity due to presence of touches of the stable and unstable manifolds for trajectories on the attractor.

Figure 5 shows a histogram of the distribution of angles between stable and unstable subspaces obtained numerically for the mentioned set of parameters for the complex amplitude equations (7). Observe that the distribution is distant from zero angles, i.e. the test confirms hyperbolicity of the attractor.

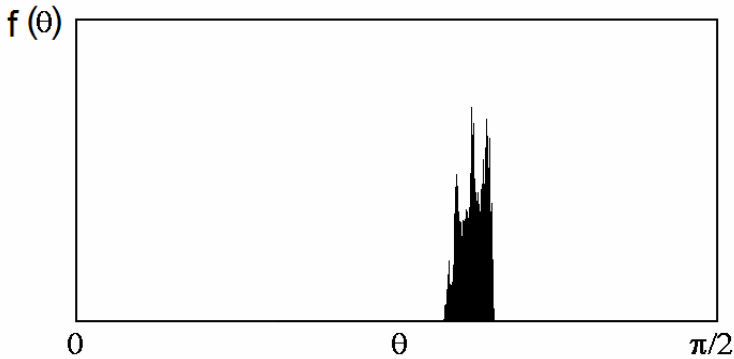

**Figure 5:** Histogram of distribution of angles of intersection of local stable and unstable manifolds on the attractor obtained from processing data of numerical iterations of the Poincaré map for the amplitude equations (7) with the procedure described in the text. Absence of near-zero angles confirms the hyperbolicity of the attractor

## 3. The circuit diagram of the parametric oscillator

For elaboration of the parametric generator of chaos as a real electronic device, it is natural to turn to a convenient and popular software product Multisim [33, 34]. Its original version called the Electronic Workbench was released in 1995 by Interactive Image Technologies Company. Since 2005, improved versions of the software are being developed by National Instruments under the name of NI Multisim. The results of the present paper were obtained with the licensed version of NI Multisim 10.1.1 purchased by Saratov Branch of IRE RAS.

Figure 6 shows a scheme composed of three oscillators represented by the oscillating circuits, one of which is formed by the inductance L1 and capacitance C1, and the second and the third, respectively, by the components L2 and C2, L3 and C3. Losses in the first and the second circuit are introduced by resistors R1 and R2. Like in the model system of the previous Section, operation of the device consists in periodic repetition of three stages. Modulation of the pump signals is provided by switches J1 and J2, which are controlled by the periodic square wave sources V4 and V6. In the first stage (the duration $T_1$=400 μs) only the key J1 is open, in the second stage ($T_2$=550 μs) both keys are closed, and in the third stage ($T_3$=50 μs) only the key J2 is open.

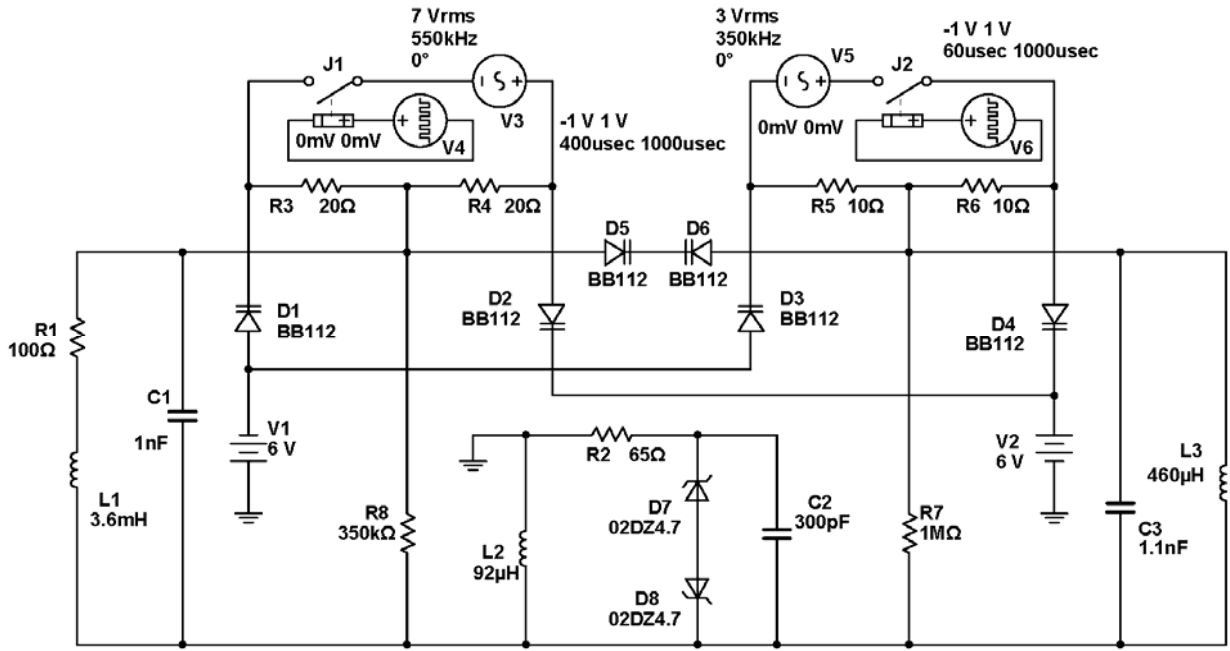

**Figure 6:** The circuit diagram of the parametric generator of chaos in Multisim
(explanations in the text)

The parametric excitation of the first and second oscillatory circuits in a first stage is provided by the element of alternating capacitance composed of a pair of varactor diodes D1 and D2 with opposite polarity in two parallel branches. The bias voltages are supplied from the DC sources V1 and V2, and the AC pump voltage is provided by the voltage divider on resistors R3 and R4 and the source V3. With such connections, dependence of the total capacity for the composite element on the applied voltage is a symmetric function containing only even terms in the power series expansion. Similarly, the element on varactor diodes D3 and D4 with bias from the same sources V1 and V2, and the alternating voltage supplied from the divider on resistors R5 and R6 and the source V5 provides pumping on the third stage of the device operation. The first and the third oscillators are connected by an element with cubic nonlinearity composed of the varactor diodes D5 and D6. Zener diodes D7 and D8 in the oscillating circuit L2, C2 introduce saturation of the amplitude of the parametric instability at a certain finite level.

Capacitances and inductances are selected in such way that frequencies of the oscillatory modes associated with primary excitation of the first, second and third LC-circuits without pumping were $f_1$=50 kHz, $f_2$=500 kHz and $f_3$=3$f_1$=150 kHz. Accordingly, the frequency of the pump in the first stage is set equal to $f_1 + f_2$=550 kHz, and in the third stage to $f_2 - f_3$=350 kHz.

### 4. Results of simulation of chaotic dynamics in Multisim

Figure 7 shows samples of time dependences for voltages on the capacitors C1, C2 and C3 obtained in simulation with the Multisim software product by means of a virtual oscilloscope connected to the respective nodes of the circuit. Roman numerals on the top of the figure indicate stages of operation of the scheme. In the first stage parametric excitation of the first and second circuits takes place; then the oscillations there decay in the second stage, while the oscillations in the third circuit persist. In the third stage the energy of the oscillations is transmitted due to the beats to the second circuit that creates initial conditions for development of the parametric instability in the next period of the pump modulation.

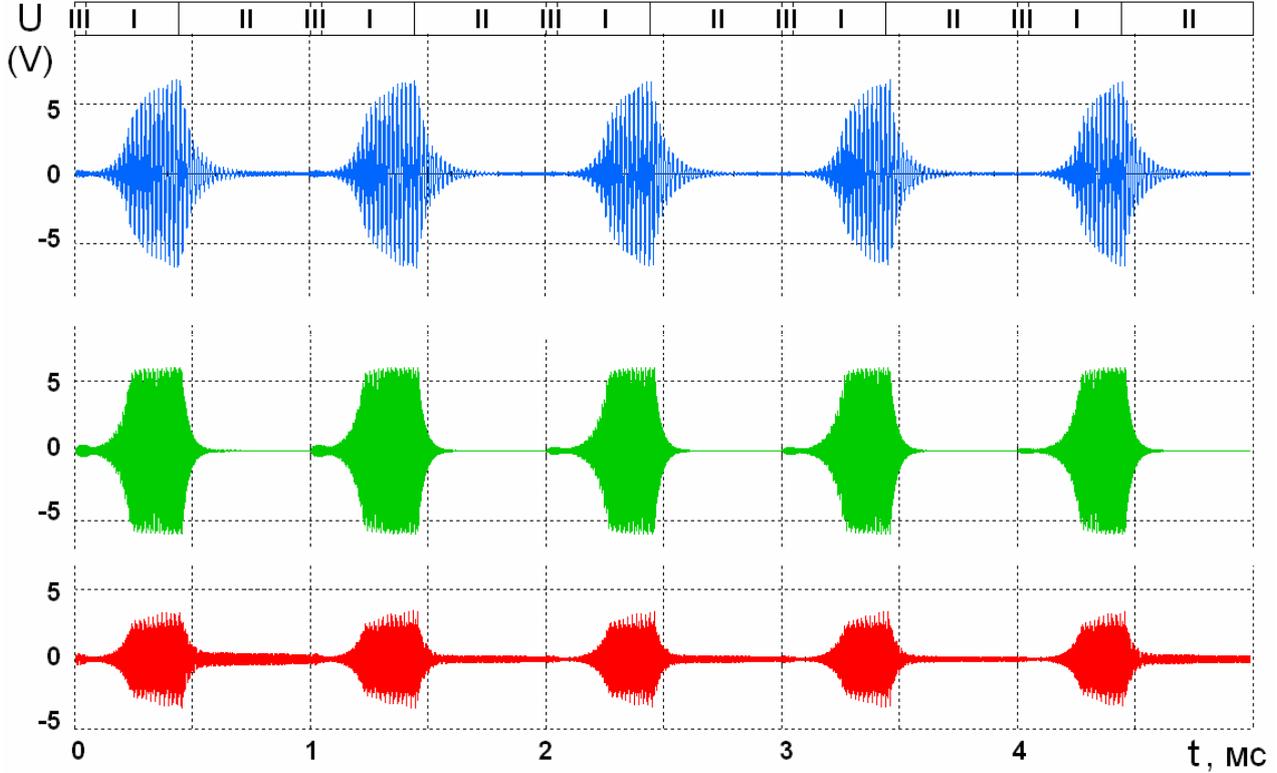

**Figure 7:** Voltages across the capacitors C1, C2 and C3 versus time copied from the screen of virtual oscilloscope during simulation of the circuit of Fig.6 in Multisim.

To be sure that passage to each new stage of the parametric excitation is accompanied with three-fold expanding map of the cyclic variable characterizing the phase of the oscillations we do the following. Let us connect the oscilloscope with the circuit in such way that one input is voltage $U_L$ across the inductance L1 and the second is the voltage $U_R$ on the resistor R1. In Multisim using the Grapher tool one can record the data in a file with a possibility of their subsequent computer processing. Time step value for the sampling has to be selected equal to the modulation period ($T=1$ ms), and the instants should correspond to epochs of intense oscillations in the first oscillator. The last condition may be controlled using delay parameters of the pulse sources V4 and V6. (In our case they were 500 and 450 μs.) Then the recorded data are processed with a specially prepared computer subroutine. First, both $U_L$ and $U_R$ are normalized in such way that the sums of squared terms become equal. Then, for each pair $U_L^n, U_R^n$ relating to one and the same time instant $t_n=nT+T/2$ we evaluate the phase as $\varphi_n = \arg(U_L^n + iU_R^n)$ defined in the interval from 0 to $2\pi$.

Figure 8 is a plot of $\varphi_{n+1}$ versus $\varphi_n$. Observe that it corresponds to the three-fold expanding map similar to that in Fig.1c. Thus, the main condition or the presence of the attractor of the Smale-Williams type is fulfilled.

The same connection of the oscilloscope can be used to produce a portrait of the attractor in the projection on the phase plane of the first oscillator. The oscilloscope is switched it to the mode in which the deflections of the beam horizontally and vertically are controlled by the input voltages $U_L$ and $U_R$. The portrait of attractor is shown in Fig. 9a. To depict the attractor in the stroboscopic section, we exploit the recorded file data used in the constructing the phase iteration diagrams and present them graphically in coordinates $U_L$ and $U_R$ using an external computer program. The stroboscopic portrait is shown in Fig. 9b. The object corresponds visually to a kind of Smale-Williams solenoid with intrinsic distinguishable characteristic structure of filaments.

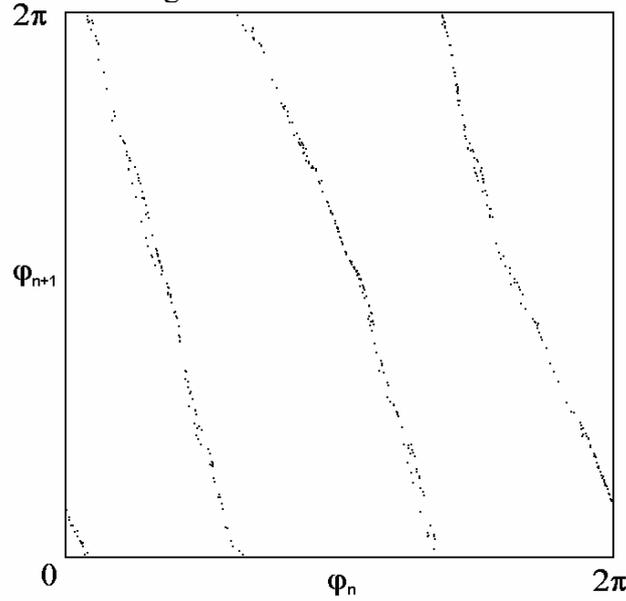

**Figure 8:** Iteration diagram for phases drawn on a base of Multisim simulation of the circuit of Fig. 6.

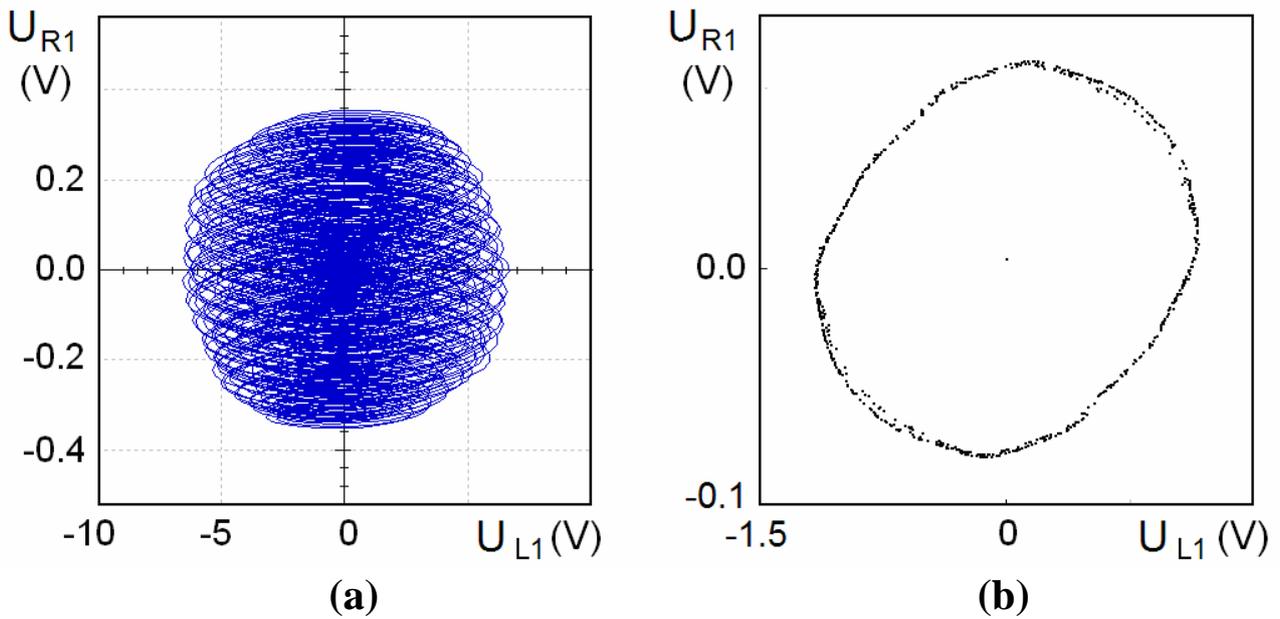

**Figure 9:** Attractor in projection from the extended phase space of the system (a) and its stroboscopic portrait (b). The horizontal and vertical coordinate axes correspond, respectively, to the voltage across the inductor L1 and the resistor R1.

Figure 10 illustrates Fourier spectra of the signals obtained using the virtual spectrum analyzer available in Multisim for the voltages across the capacitors C1, C2, C3. Spectra are continuous that reflects chaotic nature of the signals although are characterized by notable non-uniformity. For the signals from each of the three oscillatory circuits the spectra contain well-defined peaks nearby the resonance frequencies, respectively, 50 kHz for the first circuit, 500 kHz for the second, and 150 kHz for the third. Furthermore, there are peaks associated with the pump and combination frequency components due to the presence of non-linear circuit elements.

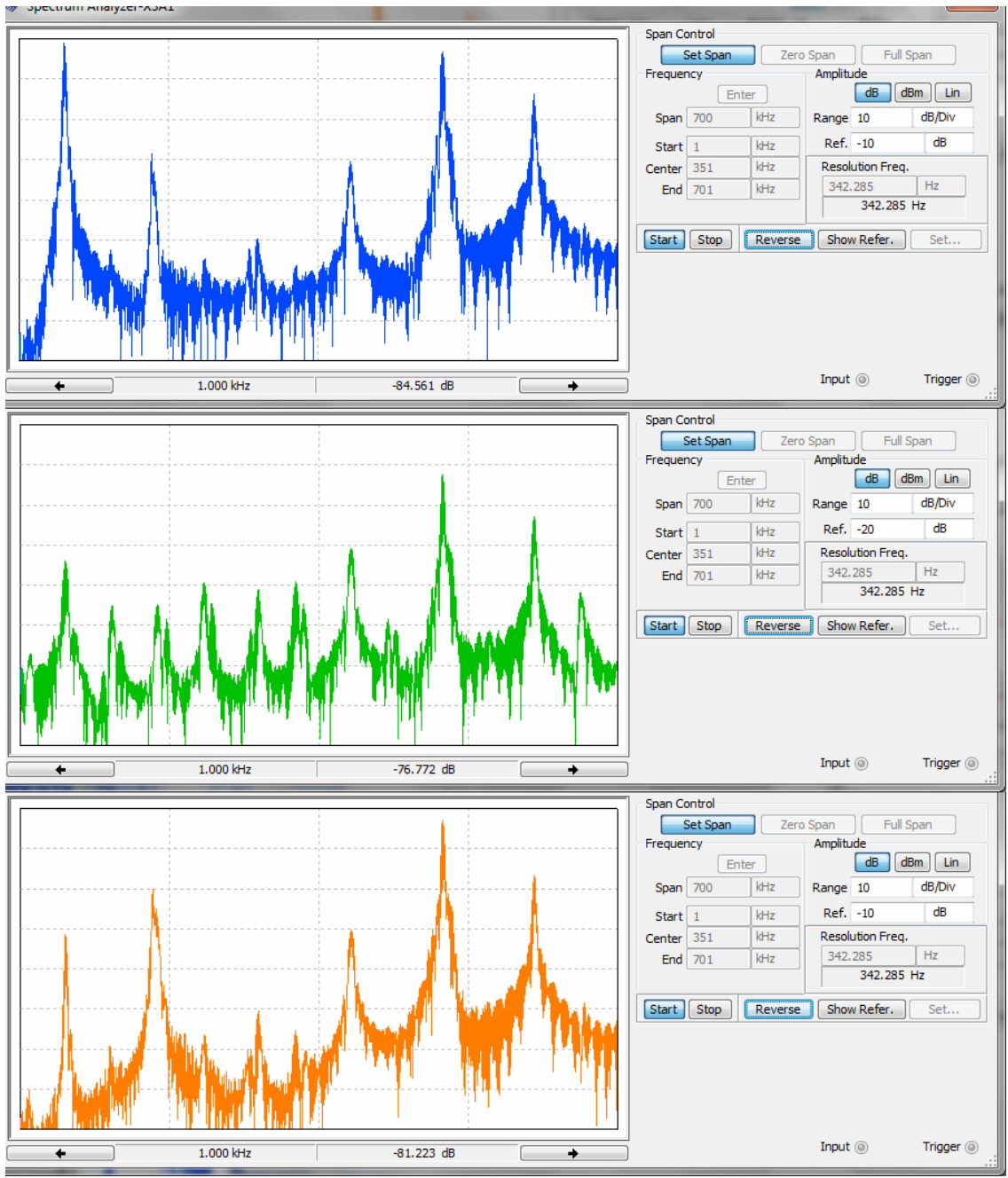

**Figure 10:** Spectra of voltage oscillations across the capacitors C1, C2, C3. The frequency range shown is from 1 to 700 kHz, the vertical axis scale is logarithmic.

**Conclusion**

In this paper, we introduce a parametric generator of chaos on a base of three coupled oscillators, in which pumping is provided by a periodic sequence of pulses of frequencies alternately equal to the sum and the difference of the natural frequencies of the partial oscillators. The system generates trains of oscillations with phases varying chaotically. The six-dimensional map describing the state transformation on a period of the pump modulation possesses a hyperbolic strange attractor, which is a variant of the Smale-Williams solenoid with tripling of the angular coordinate measured along the coils at each step of the construction. For the original flow system with such Poincaré map, according to the accepted mathematical terminology, we have an attractor in the extended phase space corresponding to a suspension over the Smale-Williams solenoid. Implementation of the system is an electronic device on varactor diodes, and its operation has been demonstrated by simulation using the software product Multisim.

The main advantage of systems with hyperbolic attractors from a practical point of view is the inherent structural stability or roughness, which means insensitivity of the properties of the generated chaos to variations of parameters and characteristics of the device components, technical fluctuations, interferences etc. In particular, such systems may be of interest as chaos generators for use in secure communications schemes [13, 35, 36]. An attractive feature is the fact that chaos reveals itself in variation of phases in the generated sequence of pulses, whereby the transmission of the signal in communication channels will be less sensitive to noise, losses and distortions comparing with other proposed schemes.

*The study was supported by RFBR under the research project № 12-02-00541.*